Author Accepted Manuscript

# Diffraction of helium and hydrogen atoms through single-layer graphene

Carina Kanitz[1], Jakob Bühler[1], Vladimír Zobač[2], Joseph J. Robinson[1], Toma Susi[2]*, Maxime Debiossac[1], and Christian Brand[1]†

[1]German Aerospace Center (DLR), Institute of Quantum Technologies; Wilhelm-Runge-Strasse 10, 89081 Ulm, Germany.

[2]University of Vienna, Faculty of Physics; Boltzmanngasse 5, 1090 Vienna, Austria.

* Corresponding author. Email: Toma.Susi@univie.ac.at

† Corresponding author. Email: Christian.Brand@dlr.de

Diffraction of atoms from surfaces provides detailed insights into structures, interactions, and dynamical processes. However, currently the method is limited to measurements in reflection – diffraction through materials has only been demonstrated for subatomic particles and is an outstanding challenge for atoms. We diffract helium and hydrogen atoms at kiloelectronvolt energies through single-layer graphene at normal incidence. Despite the atoms' high kinetic energy as well as coupling to the electronic system of graphene, we observe coherent scattering. This preservation of coherence was the result of the limited momentum transfer between the projectile and the lattice, resulting from interaction times on the femtosecond scale.

A mere seven years after de Broglie published his seminal work on the wave nature of massive particles in 1923 (*1*), diffraction had been demonstrated for electrons (*2*, *3*), atoms, and diatomic molecules (*4*). Key to this success were crystalline materials that act as gratings for matter waves. They were used in reflection for electrons by Davisson and Germer (*2*), and for atoms and molecules by Estermann and Stern (*4*). Crystal lattices also played a critical role in demonstrating diffraction of electrons in transmission by Thomson (*5*).

While electron diffraction revolutionized microscopy (*6*), diffraction and interferometry of atoms became an indispensable tool for modern physics (*7-14*). Atom interferometers are used to measure atomic properties (*15, 16*), define fundamental constants (*17-20*), and search for new physics beyond the standard model (*21, 22*). Atom diffraction from surfaces provides information on materials, such as electronic structure, binding interactions, and dynamical processes (*11, 12*). The latter includes the movement of adsorbents on the surface, ranging from the diffusion of atoms (*11, 23*) and molecules (*24, 25*) to layer deposition in molecular beam epitaxy (*26*). Atom diffraction is complemented by incoherent scattering, which is characterized by an exchange of energy between the atoms and the surface (*27*). This provides additional insights into mechanical properties and dynamics of the surfaces (*28-32*), their internal states (*33-36*), and even the existence of transient bonds (*37*). Thus, atom-based experiments combine a neutral probe with a high sensitivity to a wide variety of physical phenomena.

Given the broad success of electron diffraction in transmission, it is highly attractive to also establish atom diffraction through materials. This corresponds to the atomic counterpart of the transmission experiment done with electrons by Thomson (*5*). However, atom diffraction at surfaces has been restricted to reflection, as there are several requirements that must be met to obtain useable diffraction data. The coherence of the matter wave must be preserved during transmission to render diffraction possible, and requires that the interaction between atom and crystal is sufficiently weak to conceal the exact path of the projectile through the grating. During the passage, the distance between the matter wave and the grating atoms is on the order of 1 Ångström, leading to a substantial overlap of the atomic orbitals of the projectile with those of the grating. Transmitting an atom through a material while preserving coherence thus seems a formidable task.

We now demonstrate diffraction of helium and atomic hydrogen through free-standing single-layer graphene. We use atoms with a kinetic energy component $E_0$ normal to the lattice of up to 1.6 keV, three orders of magnitude larger than in any previous atomic diffraction experiment (*11, 38*). Diffraction in this regime is even more surprising, as it was previously shown that decoherence prevails as soon as the kinetic energy normal to the surface exceeds only a few electron volts (*12*). Furthermore, such atoms have sufficient energy to damage the crystalline grating (*39*).

## Atom-Crystal interactions

We chose single-layer graphene as a grating for atoms in transmission given its electronic and mechanical properties. Its high conductivity effectively dissipates electronic excitations, making it robust against irradiation damage. At the same time, the strong atomically-thin membrane limits coupling with the projectiles, increasing the chances for coherent transmission compared to a thin graphite sample, for instance.

It can be routinely prepared as free-standing material on suitable support structures. Regarding the matter wave, helium was a chosen as it is inert and the smallest neutral atom. Although atomic hydrogen is lighter than helium, its unpaired electron leads to stronger couplings to graphene (*37*, *40*), which might destroy the coherence of the matter wave.

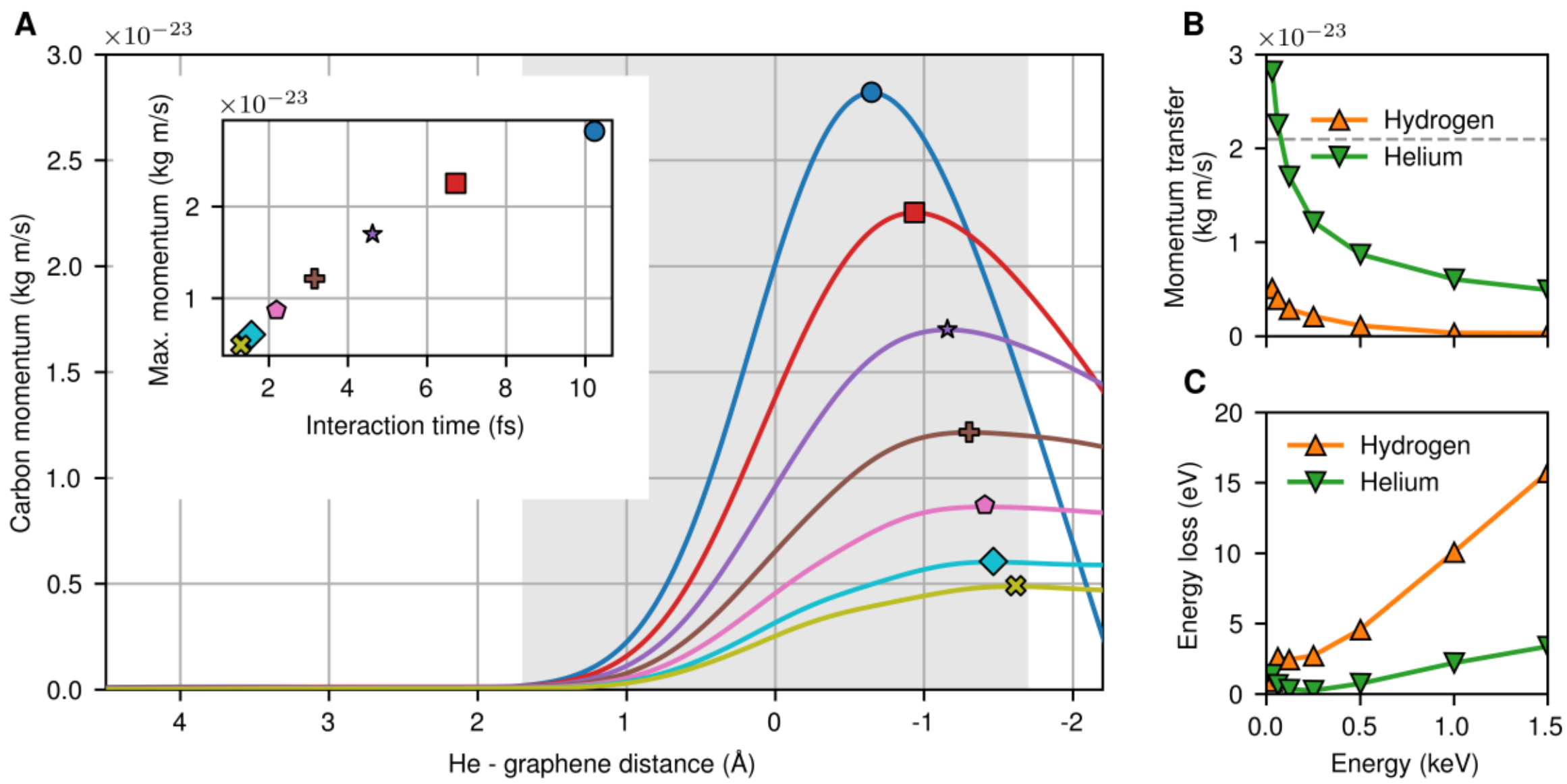

**Fig. 1 | Coupling of fast atoms to graphene**. (**A**) Transverse momentum of a carbon atom in graphene during the hexagon-center transmission of helium at different kinetic energies: 30 eV (circle), 60 eV (square), 120 eV (star), 250 eV (plus), 500 eV (pentagon), 1,000 eV (diamond), and 1,500 eV (cross). The gray area shows the van der Waals interaction region of ±1.7 Å. The inset shows the corresponding interaction time (duration for the projectile passing through the interaction region). (**B**) Transverse momentum transfer to a carbon atom while He or H transmit the lattice depending on the projectile kinetic energy. The broken line corresponds to the in-plane momentum uncertainty of C in graphene. (**C**) Energy loss to the electronic system while He or H pass the material depending on the projectile kinetic energy. In (**B-C**) the symbols are calculations and the lines are to guide the eye.

To find suitable experimental conditions for diffraction, we modeled the interaction of the atoms with the crystal using time-dependent density functional theory molecular dynamics simulations (*41*) (Fig. 1 and materials and methods). The most critical parameters are the energy loss to the grating and the exchanged transverse momentum $\Delta p$ (*42*). As long as $\Delta p$ is below Heisenberg's momentum uncertainty $p_{uncert}$ of the carbon atoms in graphene, coherence should be preserved. The in-plane momentum uncertainty can be estimated as $m_C \sqrt{\langle v^2 \rangle} = 2.1 \times 10^{-23}$ kg m/s with $\langle v^2 \rangle$ the mean-square in-plane velocity and $m_C$ the mass of the carbon atom (*42*). For helium transmitted through the center of a hexagon, the simulations predict that $\Delta p$ is smaller than $p_{uncert}$ for a beam energy $E_0$ above 100 eV (Fig. 1B).
This result by itself may appear counter-intuitive because helium does not penetrate graphene at ambient conditions (*43*). Thus, we have to increase the atom's energy to push it through the lattice. However, this greater energy does not lead to an increased momentum transfer because the interaction time is decreasing (Fig. 1A). By making the atoms travel faster, we can also mitigate the increase of $\Delta p$ when the projectile passes the crystal closer to a grating atom (*42*).

A high beam energy, however, enhances the energy loss to the electronic structure of graphene (Fig. 1C). The energy loss has a minimum around $E_0 = 250$ eV, increases with beam energy, and reaches several electron volts at $E_0 = 1{,}500$ eV. A good compromise between the energy loss and the momentum transfer is expected for beam energies between 400 and 600 eV. For hydrogen, the momentum transfer to the crystalline lattice is smaller by about an order of magnitude compared to helium (Fig. 1B). At the same time, the energy loss to graphene is much more pronounced, reaching 16 eV at a beam energy of 1,500 eV.

**Coherent diffraction in transmission**

A schematic of the experimental setup is depicted in Fig. 2A and described in more detail in the materials and methods. We prepared a beam of helium ions or protons in the energy range between 390 and 1,600 eV with an ion gun and then neutralized it using a gas-filled charge-exchange cell (*38, 44*). At these energies, we expect virtually all atoms that impinge on clean single-layer graphene to be transmitted, see Supplementary Text. The de Broglie wavelength $\lambda_{dB} = h/\sqrt{2mE_0}$ of the atoms is defined by the beam kinetic energy $E_0$, the atomic mass $m$, and Planck's constant $h$. In the considered energy range, $\lambda_{dB}$ lies between 400 and 1,800 fm. At the position of the grating, the transverse coherence amounted to at least five times the lattice constant $a$ of graphene and the minimum longitudinal coherence is about $50 \times \lambda_{dB}$, and was sufficient for multislit diffraction.

The matter wave impinges with a momentum $\hbar k_0 = h/\lambda_{dB}$ at the honeycomb lattice of graphene, with reduced Planck constant $\hbar$. During diffraction, the grating imparted transverse momentum to the matter wave in integer multiples of the reciprocal basis vectors $\hbar \boldsymbol{G}_1$ and $\hbar \boldsymbol{G}_2$. Those are proportional to the inverse of the lattice constant $a$. The diffraction angle $\theta$ is then given by the ratio of imparted momentum to forward momentum $\sin(\theta) = |\boldsymbol{G}|/k_0$, where $\boldsymbol{G} = n_1\boldsymbol{G}_1 + n_2\boldsymbol{G}_2$. Both $n_1$ and $n_2$ are integer numbers. The small lattice constant led to diffraction angles on the order of several milliradians that can be well resolved with a position-sensitive detector (Fig. 2).

In our experiment, the atomic beam had a diameter of ~ 300 µm when it reached the suspended polycrystalline graphene. Instead of individual diffraction peaks arising from a single lattice orientation, we thus expect Debye-Scherrer rings (*45*), reflecting the random orientations of the different crystal domains in the sample. We resolved patterns exhibiting more than 10 distinct rings at angles reaching up to 15 mrad (Fig. 2B).

To verify that the peaks are caused by beam diffraction, we plotted the angular positions of the Debye-Scherrer rings against the de Broglie wavelength (Fig. 2C). We observed an excellent agreement between the experimental data and the theoretical diffraction angles both for H and He. We therefore conclude that the patterns result from the coherent interaction with the lattice and are the result of diffraction.

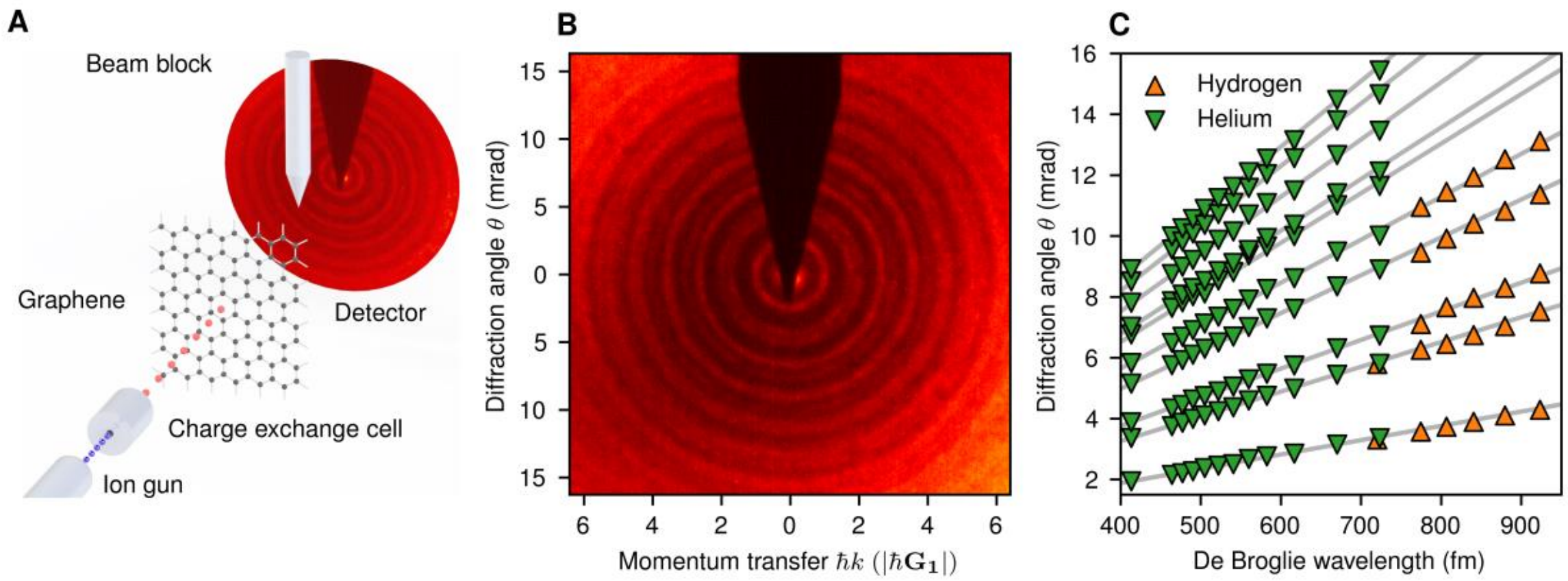

**Fig. 2 | Atomic diffraction through polycrystalline free-standing single-layer graphene**. (**A**) A beam of $H^+$ or $He^+$ (blue spheres) is prepared using an ion gun and then neutralized in a charge-exchange cell. After collimation to better than 1 mrad (full-width at half-maximum) the neutral beam (red spheres) impinges onto the graphene sample at normal incidence. The transmitted signal is visualized using a position-sensitive detector (microchannel plate) stacked onto a phosphorus screen and recorded with a camera. (**B**) Diffraction of helium at 706 eV through polycrystalline graphene results in Debye-Scherrer rings exhibiting diffraction angles of more than 15 mrad. (**C**) Experimental verification of the diffraction equation. The diffraction angle $\theta$ of H (orange) and He (green) is plotted versus the de Broglie wavelength (corresponding to kinetic energies between 390 and 1,600 eV). The lines represent expected diffraction angles $sin\,(\theta) = |\boldsymbol{G}|/k_0$.

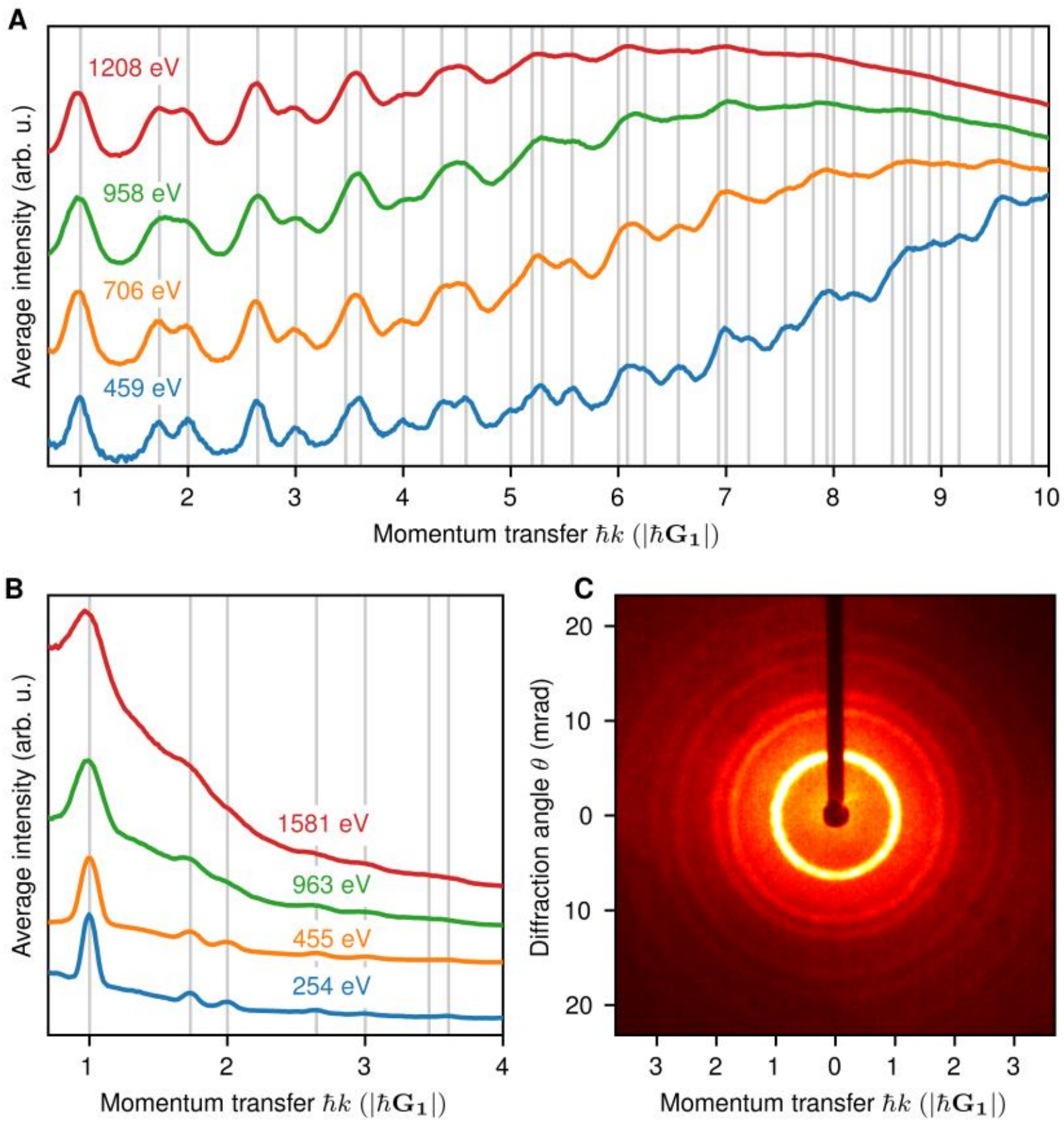

**Fig. 3 | Energy-dependent atomic diffraction through single-layer graphene**. The experimental azimuthally averaged intensity at various kinetic energies is plotted versus momentum transfer in units of the graphene reciprocal lattice vector $\boldsymbol{G}_1$ for He (**A**) and H (**B**). Vertical lines indicate the calculated diffraction angles. (**C**) Diffraction pattern of H atoms at 455 eV.

## Coupling to the crystal lattice

Whereas the beam energy determines the peak positions, simulations predicted that it also influenced the couplings to the lattice. To explore this, we compare the azimuthally averaged intensity traces for helium at various beam energies (Fig. 3A and Fig. S2). At the lowest energy, we distinguished more than 20 diffraction rings. They corresponded to about eight reciprocal vectors exchanged with the surface, as $\boldsymbol{G}_1$ and $\boldsymbol{G}_2$ span a 2D lattice (Fig. S1 and Supplementary Text). With increasing energy, the higher diffraction orders vanished and were replaced by a broad unstructured background. This background dominated at large diffraction angles, that is, when $k > 8 \cdot |\boldsymbol{G}_1|$ at $E_0 = 1{,}208$ eV. That trend persisted even when increasing the angular resolution of the experiment (see Fig. S4 and Supplementary Text). For hydrogen, the damping of higher diffraction orders was even more pronounced, and the patterns consisted only of a few diffraction orders (Fig. 3, B and C, and Fig. S3). Nevertheless, we observed diffraction for hydrogen in the energy range from 250 to 1,600 eV. These observations are fascinating, as they suggest that the atoms have lost several electron volts of energy to the grating yet still preserved their coherence. The more pronounced energy loss of hydrogen compared to helium we attributed to the unpaired electron and thus a higher reactivity, as previously observed for H at lower energies (*36*, *37*).

## Discussion and Outlook

While atomic diffraction through nanomachined gratings is a successful field since decades (*7, 46-49*), we have demonstrated diffraction of atomic matter waves through a crystal. Although we expect substantial couplings to the lattice, we resolved diffraction patterns corresponding to momentum transfers of up to $\pm 8\ \hbar|\boldsymbol{G}_1|$. To reach a comparable momentum splitting for rubidium atoms using a light grating at 780 nm, it would be necessary to scatter more than 58,000 photons, see Supplementary Text. Our platform thus realizes the beamsplitter with the largest known momentum transfer for atoms in transmission. To achieve this, we illuminated graphene at normal incidence with atomic beams in the kiloelectronvolt energy range. Surprisingly, this high energy did not lead to the grating's disintegration. Although some samples were irradiated for more than 100 hours, we did not observe any degradation of the grating's performance. As diffraction is sensitive to the phase accumulated by the interfering particles, the method provides a detailed picture about couplings of atoms with matter. At the same time, the de Broglie wavelength in the femtometer range provides an excellent resolution in momentum space.

With this study, we open the door to use complex particles for transmission experiments. In contrast to electrons, atoms exhibit internal degrees of freedom. Thus, matter waves can be easily prepared in specific internal state and thus widely expand current possibilities in transmission experiments. Such studies might have far-reaching implications for a variety of research areas. Neutral atoms in the energy regime of our experiment are omnipresent in interstellar space and the magnetosphere (*50-52*). Interference measurements might thus provide insights into interactions with surfaces or small constituents, such as polycyclic aromatic hydrocarbons. From the standpoint of materials research, it would be fascinating to extend the experiments to other crystals and their heterostructures, see Supplementary Text and Fig. S5-S7.

The use of single crystals is interesting as it would allow for separating coherent and incoherent contributions of the diffraction patterns. The method might even open new avenues to study organic and biological materials whose sensitivity to ionizing radiation remains a significant limitation (*53*). There is also a tantalizing possibility for applications in structural biology, where established cryogenic transmission electron microscopy-based techniques are limited by radiolysis damage due to the use of ionizing radiation as probe (*53*). Neutral atom beams, on the contrary, can be expected to couple less strongly to the electronic system of the target materials. This makes them a promising alternative to electrons for damage-free characterization of biological assemblies.

Finally, combining crystalline transmission gratings into interferometers might give rise to new quantum-based sensors. Fast atoms have advantages for detecting gravitational waves (*54*) compared to cold-atom experiments (*55*) and might give rise to new multi-dimensional interferometers (*56*).

**Acknowledgments:** We thank Philippe Roncin and Markus Arndt for their longstanding support of the project including many fruitful discussions and advices over the years. We further thank Maxim Efremov, Alex Wodtke, Benjamin Stickler and Jani Kotakoski for useful discussions.

**Funding:** Austrian Science Fund grant P 36264-N (VZ, TS)

**Competing interests:** Authors declare that they have no competing interests.

**Data and materials availability:** Data and simulation code used in the analysis have been deposited in Zenodo (*57, 58*). Additional materials will be made available to any researcher for purposes of reproducing or extending the analysis.

## Materials and Methods

Experimental Setup

We create a beam of protons or helium ions using a commercial ion gun including a Wien filter (Nonsequitur Technologies model 1402W). The resulting beam energy $E$ and its energy distribution $\Delta E$ are measured using a retractable Faraday cup (Kimball Physics FC-71) equipped with a retarding grid. Typical full width at half maximum values obtained for $\Delta E$ are 24 eV for $H^+$ and 18 eV for $He^+$. To neutralize the beam, the ions are guided into a 38 mm long charge-exchange cell filled with a neutral gas. We use He as neutralization gas for $He^+$ and Ar in the case of protons. The pressure inside the cell is optimized for neutral beam intensity as well as beam convergence, and is typically in the range of $5 \times 10^{-2}$ mbar. We note that neutralization of $H^+$ on Ar leads to an energy loss of 2.6 eV (*59*). However, this is too little to be observable in the experiment.
The entrance and exit of the charge-exchange cell are closed off by two pinholes that can be varied in diameter. Unless stated otherwise, the pinholes at the entrance and exit have a diameter of $s_0 = 1$ mm and $s_1 = 500$ µm respectively. Remaining ions behind the charge-exchange cell are removed from the beam with a deflection voltage. The neutral beam is collimated by a $s_2 = 200$ µm wide pinhole situated at $L = 790$ mm behind the charge-exchange cell, resulting in an angular resolution of $\approx 0.6$ mrad.
As gratings, we use commercially available monolayer graphene suspended over a holey silicon nitride substrate. These samples are positioned into the beam using a five-axis manipulator. The longitudinal coherence $\ell_l$ is given by $\lambda_{dB}^2/\Delta\lambda_{dB} \geq 50\lambda_{dB}$. The transverse coherence $\ell_t$ of the matter wave at the position of the grating can be estimated as $\ell_t = 2L\lambda_{dB}/s_1$ (*60*), resulting in $\ell_t \geq 5\,a$.
After diffraction, the atoms propagate 727 mm until they impinge on a double-stack microchannel detector stacked onto a phosphorous screen. The detector has an active diameter of 75 mm and a pore size of 10 µm. We prevent the direct beam from reaching the detector using a beam block on a linear drive. The triangular tip of the first beam block was later modified into a cylindrical beam block on a small rod to optimize the available detection region. The diffraction pattern on the detector is recorded using a CMOS camera (Hamamatsu Orca Flash 4.0) equipped with a zoom objective (Computar TEC-V7X). To reduce the background, we use a bandpass filter transmitting in the wavelength range between 535 and 558 nm (Midopt BI550) optimized for the peak emission of the phosphor P43 at 545 nm.

Graphene samples

We used commercially available monolayer graphene suspended over a holey silicon nitride substrate (Plano GmbH 21712). The sample frame is a 3 mm diameter silicon disk with a 500 µm × 500 µm window in the center (TEM grid). Over that area a patterned holey silicon nitride membrane of 200 nm thickness is suspended, which acts as a support for the free-standing graphene sample. The holes have a diameter of 2.5 µm and are positioned at a distance of 4.5 µm, constituting an open area available for transmission of 28%.
Before the start of the experiment, the graphene samples are annealed in vacuum at 450 °C for 90 min (*61*). Additionally, we clean the sample in high vacuum by illuminating it with 20 s pulses of a 300 mW cw laser emitting at 532 nm. Each pulse is followed by a 60 s cooling period and the procedure is repeated until we obtain sufficiently clean samples to observe diffraction (*62*). When diffracting H atoms through graphene, the grating was illuminated with the cleaning laser every 60 minutes to remove adsorbed atoms. In the case of helium diffraction, it was sufficient to laser-clean the sample once per day.

Molecular dynamics simulation

To simulate kinetic and electronic energy transfer from a projectile to the graphene monolayer, we employed semi-classical Ehrenfest molecular dynamics (MD) within the framework of time-dependent density functional theory (TDDFT) using the GPAW code (*63*). This enables a combined treatment of classical nuclear degrees of freedom and quantum electronic degrees of freedom. Notably, Born-Oppenheimer MD would not result in any electronic energy transfer, but this can be adequately modelled using TDDFT *(46)*.
Nuclear positions were propagated using the velocity-Verlet algorithm, while the electronic system evolution was handled by the semi-implicit Crank–Nicholson scheme with a timestep of 15 as. For a computationally efficient treatment of the required vacuum, we used localized atomic-orbital basis sets that were recently implemented for Ehrenfest MD *(45)*. We use a double-zeta polarized (*dzp*) basis for graphene and He, single-zeta (*sz*) for the H projectile, and a grid sampling of 0.2 Å with the PBE exchange-correlation functional (*64*).
Neutral H and He atoms were initially positioned sufficiently far (6 Å) above a 6×6 supercell of graphene. The projectile was then given a kinetic energy in the direction normal to the graphene plane towards the impact point, located at the center of a graphene hexagon. The momentum transfer was determined by summing the momentum components in the lateral directions of the nearest carbon atoms adjacent to the impact site. The electronic energy transfer is inferred from the loss of the kinetic energy of the projectile *(46)*, see Fig. 1.

Image correction and post-processing

Images were recorded in series. Each element of a series corresponds to 300 images with an exposure time of 1 s, which were integrated into one image. In total, we acquired data for one to four hours to arrive at the traces shown in Fig. 3. For the data needed to confirm the diffraction angles with varying energy in Fig. 2C, an acquisition time of 15 minutes was sufficient. All recorded images were compensated for radial objective distortion up to the quadratic term. Due to the software used, the intensity values needed to be rescaled before they could be combined or compared. Rescaling was performed to a region of known brightness common to all images outside of the detector area. Camera dark count correction was performed with the camera cap on with a 5-minute integration time.

After corrections, all recorded images from a measurement series were averaged together into a single image and cropped around the region of interest. To determine the center of the pattern, the inner diffraction rings were fitted with a 2D-function that was Gaussian in the radial direction and uniform along the azimuthal angle. For generating radial traces, the area of the screen covered by the beam block was excluded.

## Supplementary Text

### Transmission coefficient

The minimum kinetic energy of H (He) used in the presented experiments is 254 eV (459 eV). At these energies, we expect virtually all atoms to be transmitted through clean single-layer graphene. This has been explicitly simulated for H atoms with energies up to 200 eV using molecular dynamics (MD) simulations within a self-consistent charge density functional tight binding formalism (*65*). The authors found that the transmission probability increases with energy, reaching a value very close to 1 for H at 200 eV. For helium, to our knowledge no predictions of the transmission probability at relevant energies have been published so far. We therefore simulated the irradiation of graphene with helium and hydrogen at 1 keV, covering the unique impact points of a carbon hexagon. The minimum considered distance between the projectile and a carbon atom was 24 pm (as head-on collisions at this energy cannot be accurately modeled). Even at this close distance, the simulations predict that a hydrogen atom is transmitted and deflected only by 6.5° from a straight line. As impact parameters further away from the carbon atom also lead to transmission, the projectile can pass single-layer graphene through at least 93% of the hexagons' surface. Repeating the simulation for He at this point of impact, we observe that helium is transmitted as well and that the classical deflection angle is reduced to 3.8°.

### Debye-Scherrer rings

The radii of the Debye-Scherrer rings follow the expected positions according to the reciprocal lattice defined by the sum of $n_1\boldsymbol{G}_1 + n_2\boldsymbol{G}_2$. The magnitude of both reciprocal basis vectors $\boldsymbol{G}_1$ and $\boldsymbol{G}_2$ is $4\pi/(\sqrt{3}\,a)$ with $a = 246$ pm the lattice constant of graphene. We can visualize the positions of the rings in reciprocal space, as shown in Fig. S1. Here, the black spots represent the locations of the diffraction spots for a single crystal. The circles are the Debye-Scherrer rings resulting from azimuthally rotating each set of spots around the origin, which originate from the polycrystalline nature of the graphene samples used.

**Fig. S1.**

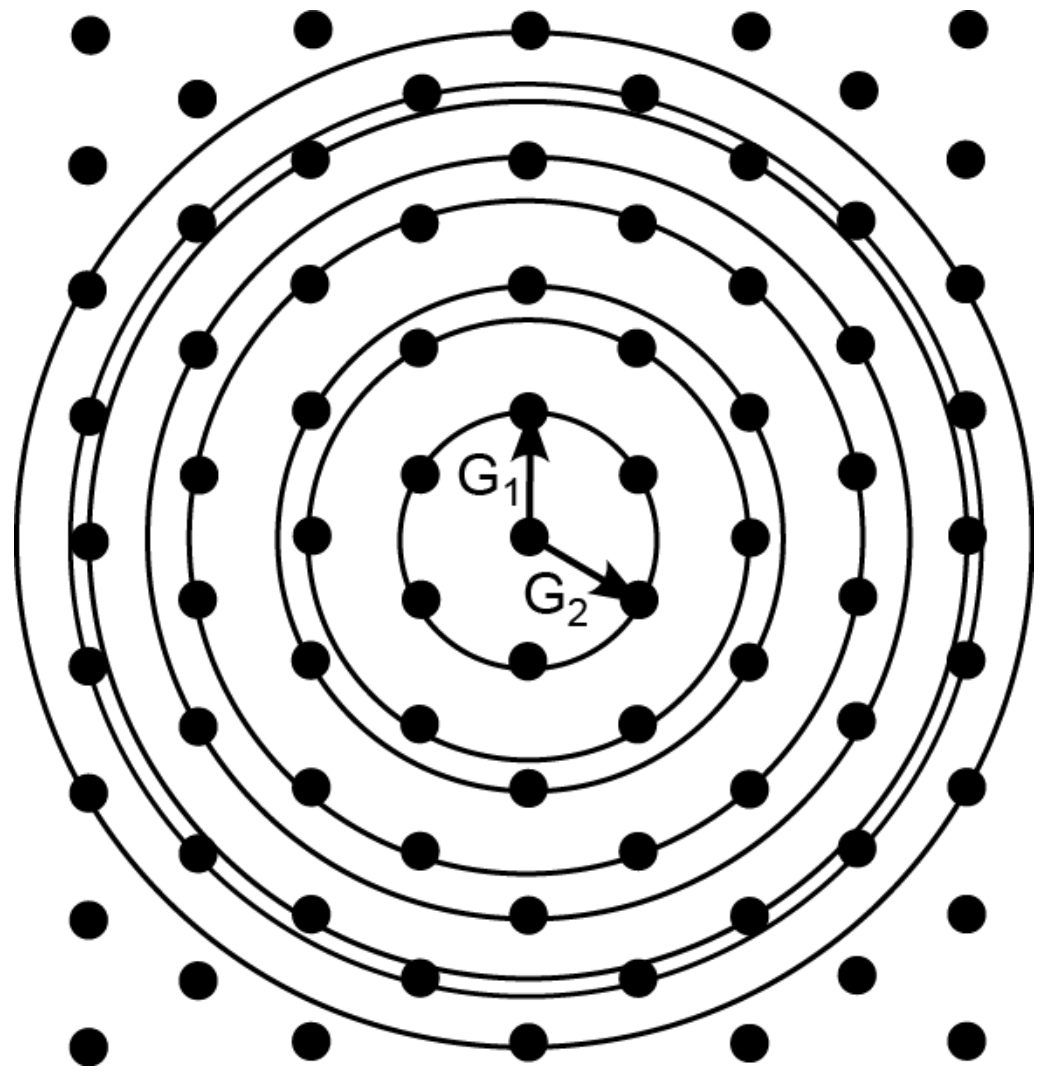

***Debye-Scherrer rings corresponding to the diffraction spots for monolayer graphene.*** *The vectors* $\boldsymbol{G}_1$ *and* $\boldsymbol{G}_2$ *are the reciprocal vectors of the lattice.*

<u>Diffraction patterns</u>

The diffraction patterns for atoms through single-layer graphene for different kinetic energies are shown for He in Fig. S2 and for H in Fig. S3.

**Fig. S2.**

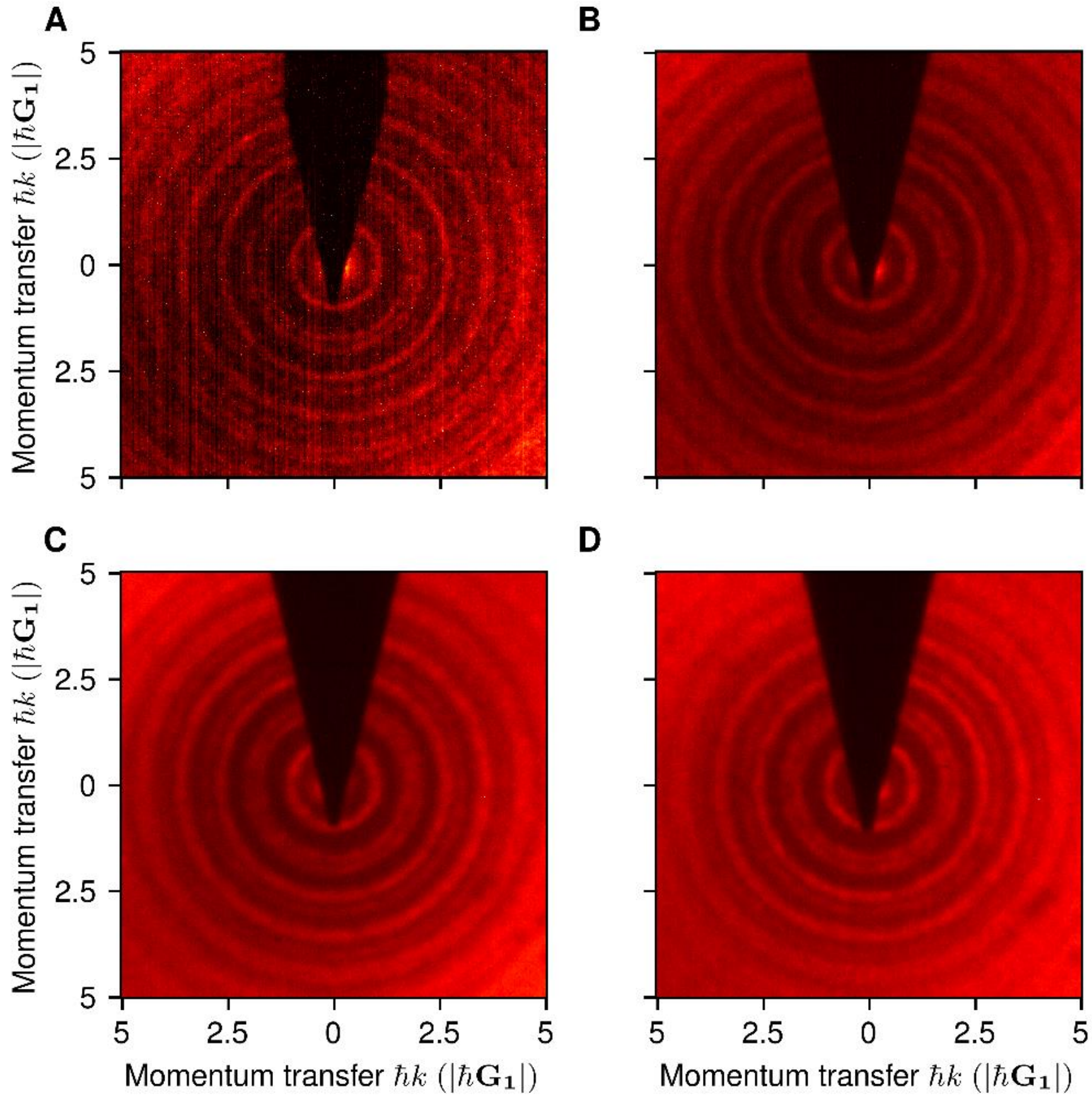

**Diffraction patterns for He passing through single-layer graphene corresponding to the profiles shown in Fig. 3A**. *Kinetic energy of the He atoms:* (**A**) *459 eV*, (**B**) *706 eV*, (**C**) *958 eV*, (**D**) *1,208 eV. The black triangular shape is the shadow of the beam block on the detector.*

**Fig. S3.**

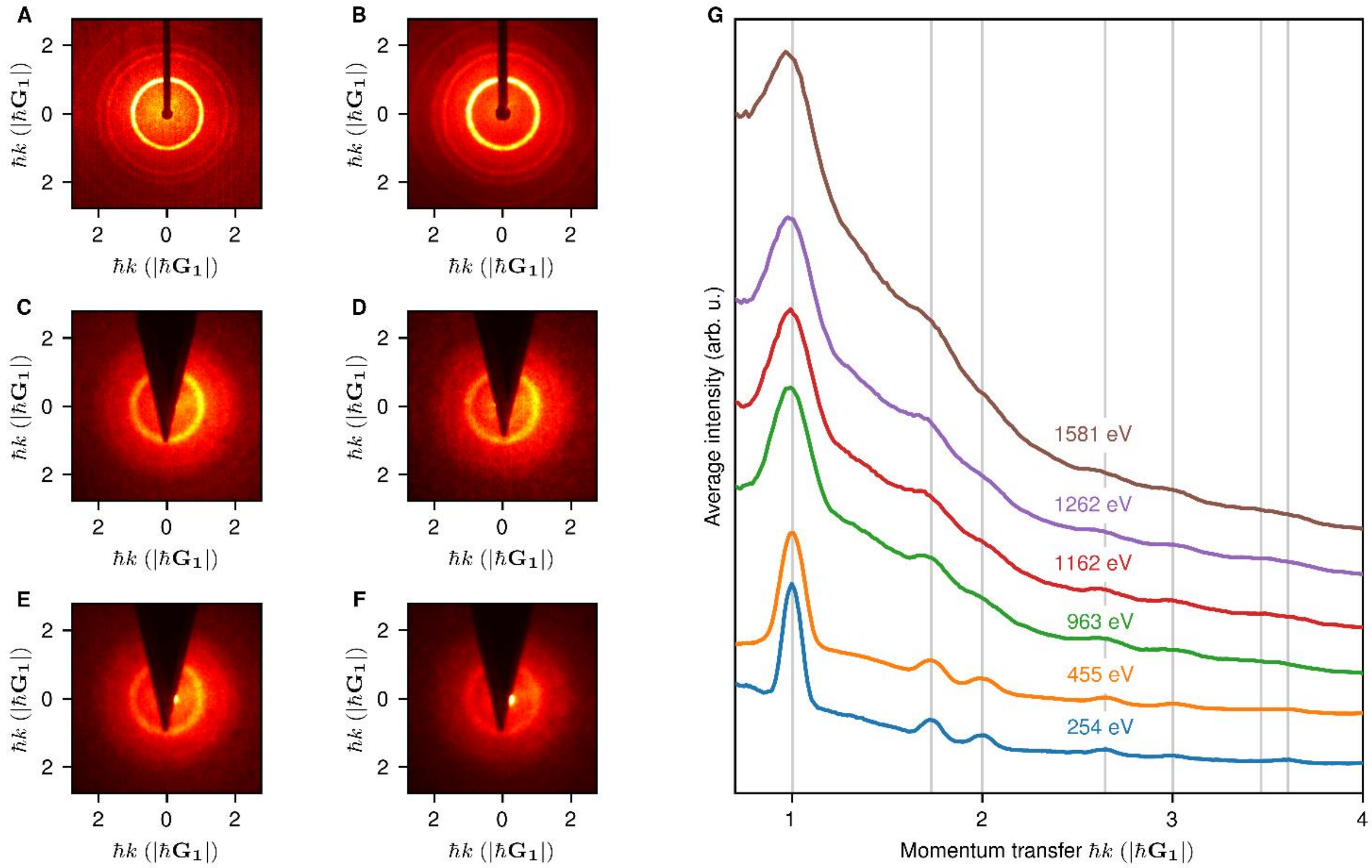

**Diffraction patterns for hydrogen passing through single-layer graphene corresponding to the profiles shown in Fig. 3B plus two additional energies**. *Kinetic energy of the H atoms:* (**A**) *245 eV,* (**B**) *455 eV,* (**C**) *963 eV,* (**D**) *1,162 eV,* (**E**) *1,262 eV,* (**F**) *1,581 eV. The shadow of the beam block on the detector is shown by a black circular shape in panel* (**A**) *and* (**B**) *and by a black triangular shape in panels* (**C**) *to* (**F**). *The azimuthally averaged intensity traces for all energies are plotted versus momentum transfer in units of the graphene reciprocal lattice vector* $\boldsymbol{G}_1$ *on the right* (**G**).

Angular resolution of the diffraction pattern

The traces depicted in Fig. 3A have been recorded at an angular resolution of 0.6 mrad. As the spacing of the diffraction orders decreases with increasing beam energy $E_0$, also the effective angular resolution decreases with increasing $E_0$. To test whether this effect explains the observed loss of visibility for higher diffraction orders, we increased the angular resolution by a factor of 1.5 by decreasing the size of pinhole $s_1$ to 200 µm. This way we can compensate for the decrease in diffraction angle by a factor of 1.6 when increasing $E_0$ from 460 to 1,208 eV. As shown in Fig. S4, the level of detail remains comparable, especially for higher diffraction angles. Thus, we conclude that the effect of collimation cannot account for the observed loss of visibility at larger diffraction angles.

**Fig. S4.**

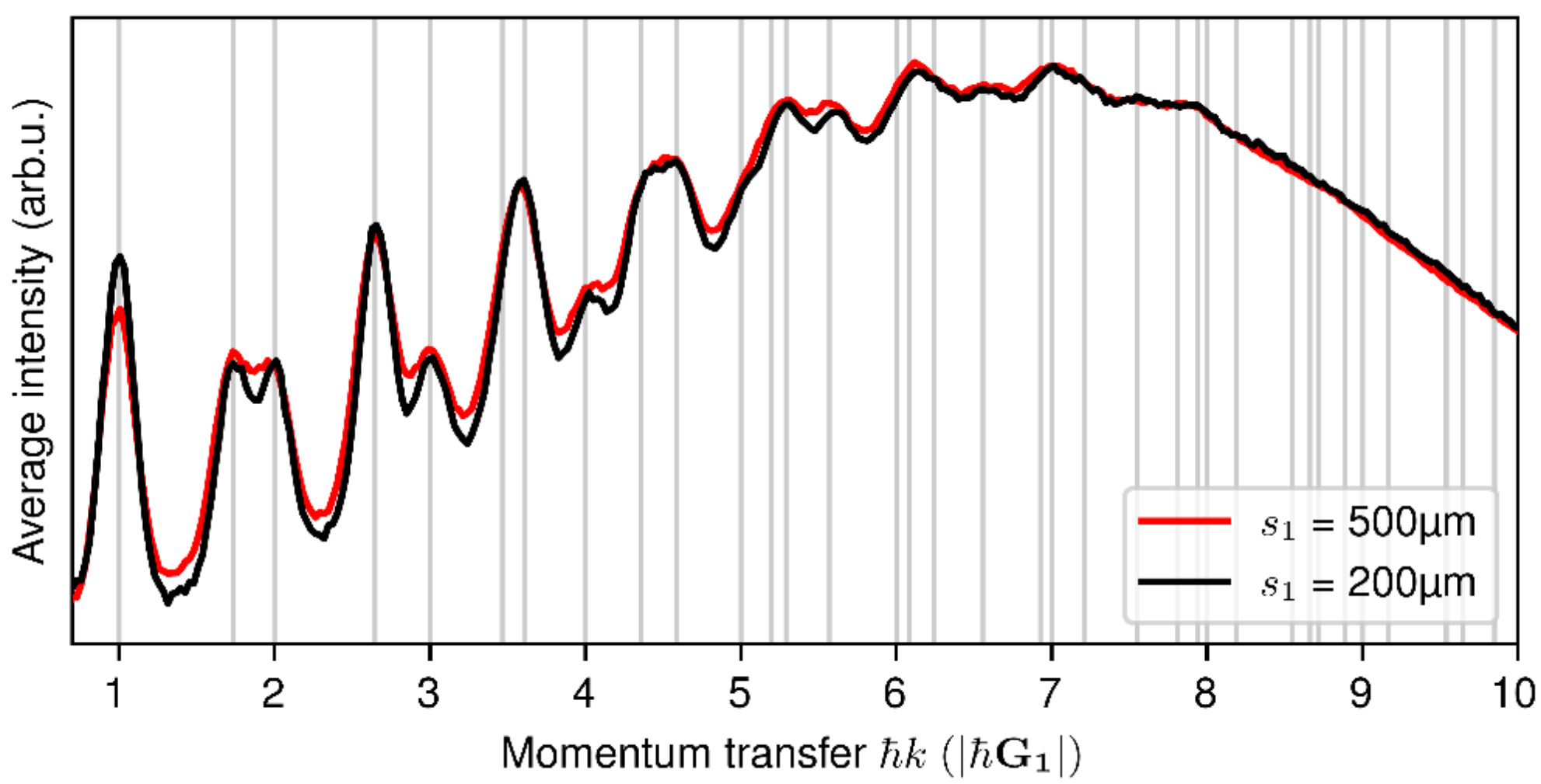

**Azimuthally averaged intensity for helium at 1,208 eV**. *Increasing the angular resolution from 0.6 mrad (red curve) to 0.4 mrad (black curve) by decreasing pinhole* $s_1$ *to 200 µm has little effect on the pattern. In both cases, we observe no structured features for a scattering of more than* $8\,\hbar|\boldsymbol{G}_1|$.

Large momentum transfer beamsplitter

There is considerable interest in creating large momentum transfer beamsplitters to increase the sensitivity of matter-wave interferometers (*66, 67*). In this context, typically laser-based gratings are used. In combination with accelerated optical lattices (Bloch oscillations) these can transfer up to a thousand photon momenta ($\hbar k$) *(17)*. We can compare the momentum transfer obtained by such an optical grating at a wavelength of λ = 780 nm used to manipulate rubidium atoms to the one obtained by diffraction at a lattice (n $\hbar \boldsymbol{G}_1$). The magnitude of the grating vector is given by $|\boldsymbol{G}_1| = 4\pi/(\sqrt{3}\ a)$ and depends on graphene's lattice constant a = 246 pm. This is larger by a factor of 3,660 than the wave vector k = 2π/λ of a photon at λ = 780 nm. Thus, to achieve the same momentum separation between two diffraction orders separated by 16 $\hbar\boldsymbol{G}_1$ (±8 $\hbar\boldsymbol{G}_1$), scattering of 16 × 3,660 = 58,560 photons would be required. Therefore, we realize the beamsplitter with the currently largest momentum transfer in transmission.

Attenuation of atomic diffraction and prospects for other materials

Atomic diffraction in transmission constitutes a novel neutral probe, making it a promising basis for materials research more broadly, especially if sufficient coherence can be retained for other specimens. An obvious field of interest is two-dimensional materials, which are expected to encompass more than 1,800 different constituents (*68*), and whose van der Waals heterostructures have attracted tremendous recent interest (*69*). Besides naturally occurring 2D materials and their heterostructures, also artificial membranes, such as covalent organic frameworks (*70, 71*) might be interesting fields of research. In this case, transmission through such membranes could allow for studying the interaction inside the pores.

To evaluate the possibility to extend our method also to multi-layered materials, we simulated the transmission of He and H atoms at 1 keV through bilayer graphene. The considered points of impact correspond to the sites with the highest probability for transmission and are shown in Fig. S5. As in the case of single-layer graphene, the critical parameters are the in-plane momentum transfer and the energy-loss of the particle.

The results for helium transmitted through single-layer graphene as well as AA- and AB-stacked bilayer graphene are shown in Fig. S6. The respective results for hydrogen are depicted in Fig. S7. The simulations indicate that the energy loss and momentum transfer to the first layer of a stacked material are nearly identical to those of a single layer. The coupling to the second layer, however, is more complex. For helium, the energy loss is slightly more than doubled compared to a single layer. For AA-stacked graphene, it amounts to 2.3 eV (single) and 5.6 eV (double) while it increases to 13.2 eV (single) and 28.1 eV (double) for an AB-stack. For hydrogen, on the other hand, the energy loss is slightly less than doubled when going from one to two layers (8.1 vs. 15.9 eV for AA, and 14.8 vs. 25.9 eV for AB).

Regarding the momentum transfer, the effect depends on the stacking. For AB-stacking, the transfer from both projectiles to the second layer is roughly 50% higher than the one to the first layer (Figs. S6D and S7D). The reason for this is that the interaction in the first layer deflects the projectile towards the C atom in the second. However, even though the momentum transfer for helium is higher than the momentum uncertainty $p_{uncert}$ of carbon in graphene (cf. Fig S6D), the probability for elastic diffraction predicted by the Debye-Waller factor is still at a few percent. In the case of AA-stacking, the momentum transfer for He strongly resembles the single-layer situation and is well below $p_{uncert}$, see Fig. S6B. For both stacking orders, the momentum transfer for hydrogen (Fig. S7) is at least an order of magnitude below the in-plane momentum uncertainty of C atoms in graphene, suggesting that coherent transmission is very likely.

**Fig. S5.**

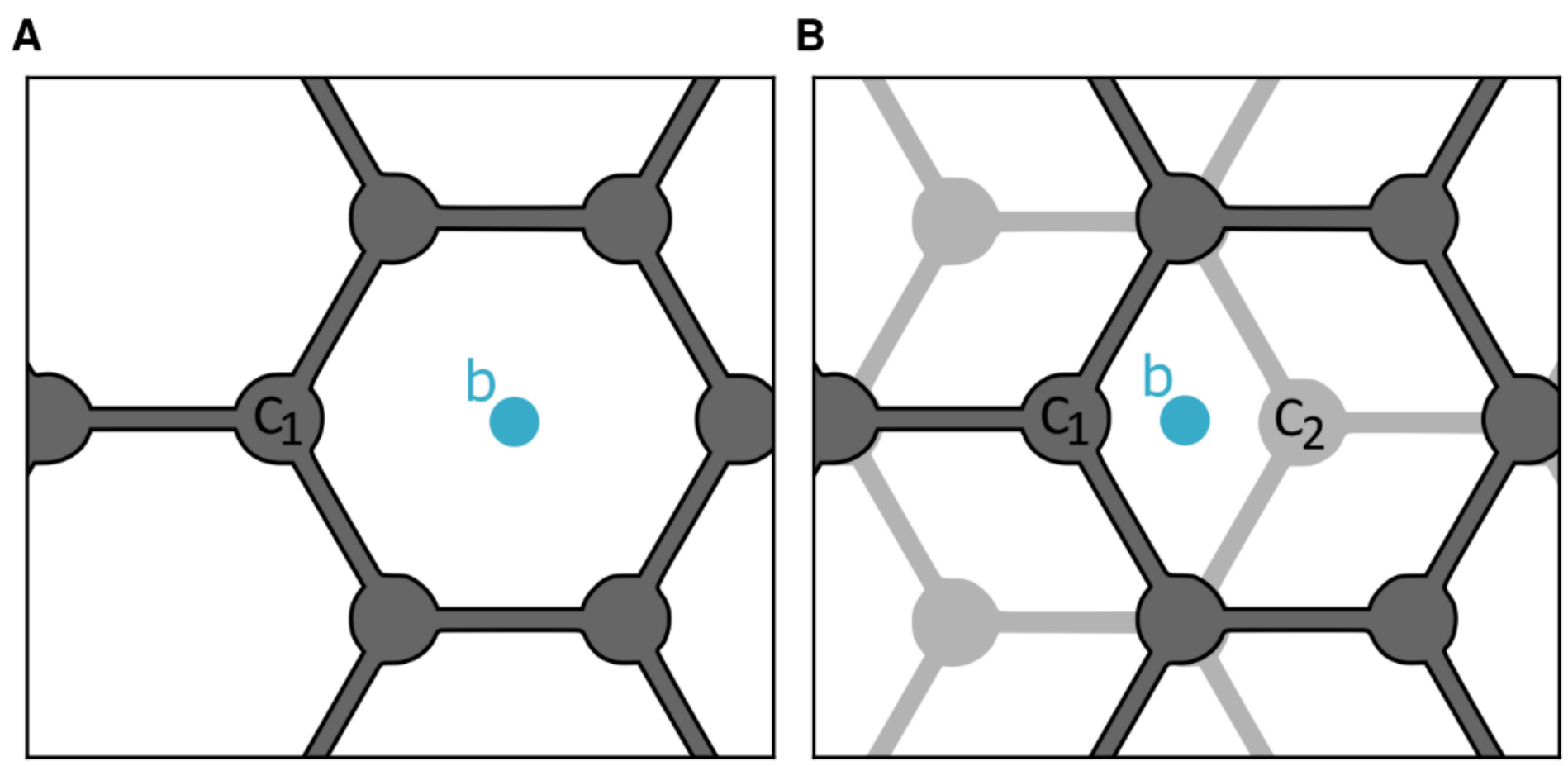

**Impact parameter b for transmission simulations for AA- and AB-stacked graphene**. *For AA stacked graphene* (**A**) *the atoms of the second layer lay exactly behind the first. For AB stacked graphene* (**B**) $C_1$ *and* $C_2$ *are the atoms the projectile has the closest approach to during transmission.*

Overall, these results suggest that while a second layer will result in greater attenuation, the magnitude is low enough that we do expect coherent transmission to remain possible, provided the stacking order allows a sufficiently large open area to be retained. Experimental realizations may, however, be hampered by the considerable challenges to prepare large-scale bilayer graphene in a single stacking order. In this context we note that AB-stacking for graphene is a consequence of the specifics of its van der Waals interlayer interaction, whereas other 2D materials prefer different stacking orders (AA' with open hexagons for bilayer hexagonal boron nitride, for example).

**Fig. S6.**

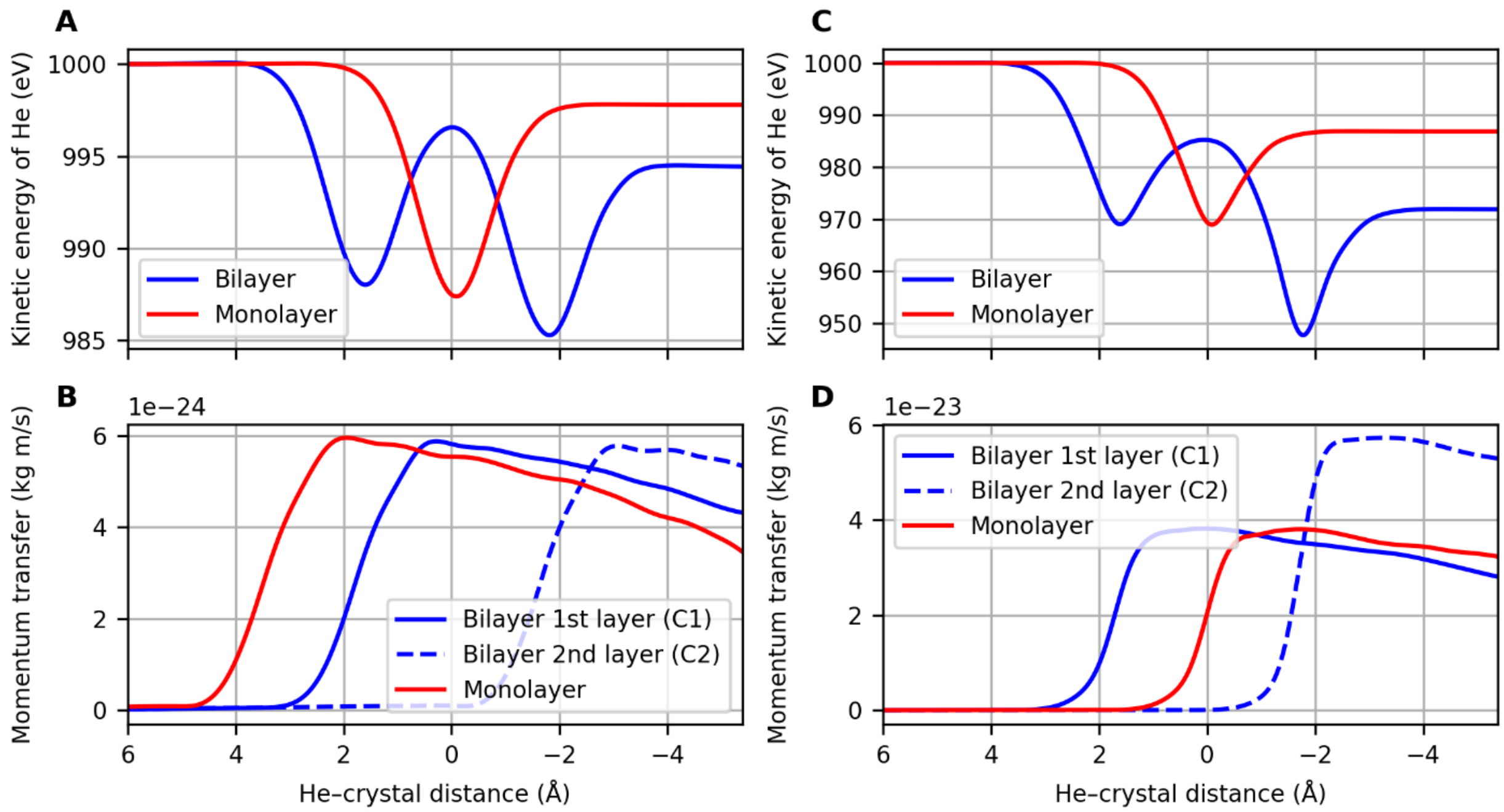

**Coupling of 1 keV He to AA- and AB-stacked bilayer graphene**. *Energy loss* (**A**) *and in-plane momentum transfer* (**B**) *during transmission through the hexagon-center impact point of AA-stacked bilayer graphene (cf. Fig. S5A). The respective energy loss and momentum transfer for the transmission through the high-symmetry impact point of AB-stacked bilayer (cf. Fig. S5B) are shown in* (**C**) *and* (**D**)*. All data have been simulated using Ehrenfest dynamics.*

**Fig. S7.**

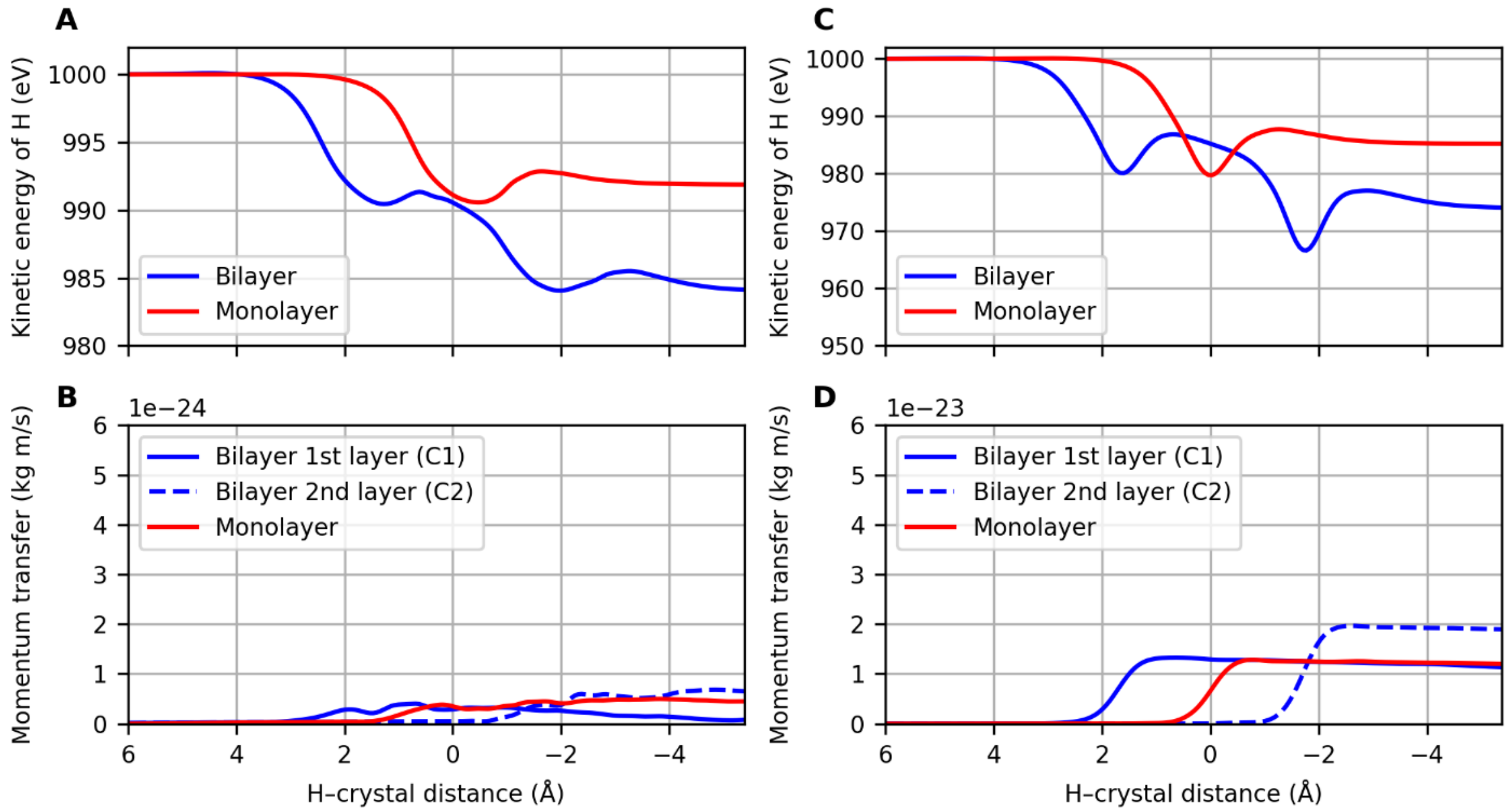

**Coupling of 1 keV H to AA- and AB-stacked bilayer graphene**. *Energy loss* (**A**) *and in-plane momentum transfer* (**B**) *during transmission through the hexagon-center impact point of AA-stacked bilayer graphene (cf. Fig. S5A). The respective energy loss and momentum transfer for the transmission through the high-symmetry impact point of AB-stacked bilayer (cf. Fig. S5B) are shown in* (**C**) *and* (**D**)*. All data have been simulated using Ehrenfest dynamics.*